\documentclass[11pt]{article}
\usepackage[margin=1in]{geometry}
\usepackage{amsmath,amssymb}
\usepackage{graphicx}
\usepackage{booktabs}
\usepackage{array}
\usepackage[authoryear,round]{natbib}
\usepackage{xcolor}
\usepackage{caption}
\usepackage{microtype}
\usepackage{hyperref}
\hypersetup{hidelinks,pdftitle={Correcting for Differential Nonresponse Using Past Election Data},pdfauthor={David Metrick, Harper Wang}}
\usepackage{float}
\usepackage[section]{placeins}

\newcommand{\logit}{\operatorname{logit}}

\title{\textbf{Correcting for Differential Nonresponse Using Past Election Data}}
\author{David Metrick, Harper Wang\\ Columbia University \\ dm4015@columbia.edu, jw4645@columbia.edu}
\date{November 2026}

\begin{document}
\maketitle

\begin{abstract}
In an election survey, the probability of responding can depend on the vote the survey is trying to measure. When supporters of one candidate are less likely to respond than supporters of the other, they are underrepresented in the sample, causing the estimated vote share to differ systematically from the true share in the electorate. Standard weighting and multilevel regression and poststratification (MRP) adjust for differences in observed covariates between the sample and the population, but they do not eliminate this bias when response rates still differ by vote within demographic groups. We ask whether this differential nonresponse can be corrected using the previous election, for which the outcome is known. We estimate a two-parameter correction on the log-odds scale against the results of one election and apply it, without modification, to the next. The corrected estimates are then compared with the next election's results. The application is the Cooperative Election Study for the 2020 and 2024 U.S.\ presidential elections, with the 436 congressional districts as units. Among validated voters in 2020, Republican voters appear in the sample at about 0.6 times the odds of Democratic voters, and in both elections a district-level MRP estimate understates the Republican two-party share by 16 to 18 percentage points on average. A correction estimated on 2020 and applied to 2024 gives 95\% posterior predictive bands that contain the actual result in 98.2\% of districts, compared with 91.5\% for a correction applied on the share scale. Approximately two-thirds of the curvature observed in 2024 was already present in the 2020 error structure, and the transferred correction removes about 69\% of the 2024 curvature; the remaining third of the curvature, a level shift of 1.7 points, and a few districts (several with very large 2020-to-2024 swings) are specific to 2024. We also fit a model for variation in differential nonresponse across districts  \citep{little1998modeling} and find that the survey's design weights account for about 62\% of the selection on the log-odds scale. Two variants that apply the correction within poststratification cells were specified in advance and did not improve on the simpler correction.
\end{abstract}

\section{Introduction}

Differential nonresponse occurs when the probability that a person responds to a survey depends on the quantity the survey is measuring. In an election survey, supporters of one candidate may be less likely to respond than supporters of the other. Weighting and multilevel regression and poststratification (MRP) adjust the sample to match the population on demographic and geographic variables \citep{gelman1997poststratification,park2004bayesian,ghitza2013deep}. Neither adjusts for the outcome itself, so differential nonresponse within demographic cells passes through to the estimates. \citet{little1998modeling} model it directly, with a group-level relative response rate for the supporters of each candidate. Differential nonresponse is now thought to account for much of what was once read as opinion change in campaign polls \citep{gelman2016mythical}. The total error of pre-election polls is about twice what their reported margins of error imply, the excess being non-sampling error of which differential nonresponse is one component \citep{shiranimehr2018disentangling}. In 2016, polling error varied by place, and the swings that decided the outcome were concentrated in particular groups even though departures from a national swing were small in aggregate \citep{gelman2017nineteen}.

A correction for differential nonresponse can only be estimated where the outcome is known, and elections provide this. Results are published for every geographic unit, so the bias of a survey estimate can be measured directly, a correction can be estimated against one election, and the correction can be applied to the next election in an attempt to improve prediction. The question is whether the bias is stable enough from one election to the next for the transfer to work, and what the residual after the transfer shows. \citet{meng2018statistical} decomposes the error of a survey estimate into the correlation between responding and the outcome, the sampling fraction, and the variability of the outcome, and \citet{isakov2020towards} use estimates of that correlation from 2016 to correct 2020 state polls.

The Cooperative Election Study (CES) is a good setting for this test. It is the largest academic election survey, with about 60,000 respondents per cycle, and it is the standard source for subnational estimates of opinion \citep{lax2009how,caughey2019public}. Its respondents are validated against voter files, and its 2020 and 2024 waves place a median of 62 and 50 validated voters in each of the 436 congressional districts, respectively, which span the full range of partisanship. \citet{chooi2026persistent} analyzes the 2024 CES at the state level with Meng's decomposition and finds Trump voters under-represented even after sample matching and vote validation, effective sample sizes of 12 to 22 per state regardless of the nominal sample, and a prior-election correction that halves state-level error. We use districts as the units and condition on demographics through MRP, so the correction we estimate is net of sample composition.

This paper makes four contributions. First, we develop and evaluate a two-parameter correction on the log-odds scale, a shift and a stretch, against the 2020 results, and apply it without modification to 2024. The corrected 95\% bands contain the actual 2024 result in 98.2\% of districts, compared with 91.5\% for a correction applied in the share scale in the style of \citet{heidemanns2020updated}. Second, we characterize the residual after the transfer and decompose it into a part that is the same in both elections, about two-thirds of the curvature in the uncorrected estimate's error, and a part specific to 2024: the other third, a level shift of 1.7 points, and a few districts outside the corrected bands. Third, we fit a model
for variation in
differential
nonresponse across
districts \citep{little1998modeling} and estimate the relative response rate three ways: from the counts alone, from the design weights alone, and from the counts given the weights. The three estimates agree on the log-odds scale, and they imply that the design weights account for about 62\% of the selection. Fourth, we specify in advance and test two variants that apply the correction within poststratification cells. Neither improves on the simpler correction, and the reasons they fail show which features of the residual do not transfer between elections.

The paper is organized around a single transfer design: the correction is estimated on the 2020 election, applied without modification to 2024, and judged out of sample. Section 2 describes the two CES cycles and the poststratification frame, emphasizing the two choices that make the pipeline usable for prediction---weighting on pre-election information only, and treating the analytic sample as the product of a composite selection screen rather than panel recruitment alone. Section 3 gives the methods in the order they are used: the MRP and poststratification that produce the uncorrected estimates (3.1); the two corrections and how their parameters are estimated on the earlier election (3.2); the validation procedure (3.3); a model of district-level variation in differential nonresponse, used both as a direct estimator and as a benchmark (3.4); and two cell-level variants specified in advance (3.5). Section 4 reports results: the level and shape of the uncorrected bias (4.1), the out-of-sample transfer and its calibration (4.2), the residual structure that remains (4.3), a decomposition of selection into weighting and residual components (4.4), and the failure of the cell-level variants under pre-specified promotion criteria (4.5). Section 5 discusses which features of the bias are stable properties of the survey and which are specific to the 2024 electorate.

Every number reported comes from a seeded run of the complete pipeline; Appendix~\ref{app:repro} describes the checks that were needed to make the pipeline reproducible.

\section{Data}

\subsection{Surveys}
We use the 2020 and 2024 CES Common Content files \citep{ces2020,ces2024}, restricted to respondents whose general-election turnout was validated against the voter file (Catalist for 2020, TargetSmart for 2024) and who reported a vote for one of the two major-party presidential candidates. The outcome is an indicator for a Republican presidential vote. After listwise deletion on the four demographic variables, the analytic samples contain 26,951 respondents (2020) and 22,061 (2024). Districts have a median of 62 respondents in 2020 (range 19--131) and 50 in 2024 (range 12--94); all 436 districts, including the District of Columbia, are represented in both years.

In this paper, we make two key choices with respect to the survey data. First, we treat the CES design weight (\texttt{commonweight}, restricted to validated voters) as the weight variable in both years, rather than the voter-validated weights supplied for 2020 and 2024. We do this because, in order to predict future elections, we would not have access to after-the-fact voter-validation weighting, so we need weighting that uses only pre-election information. Second, respondents enter the analytic sample through a composite screen---panel membership, voter validation, and answering the presidential item---so the selection quantities estimated below describe that composite, not panel recruitment alone.

\subsection{Poststratification frame}
The poststratification frame has 120 cells per congressional district: age (18--29, 30--44, 45--64, 65+), gender (male, not male), race/ethnicity (non-Hispanic White, non-Hispanic Black, Hispanic), and education (no high school diploma, high school, some college, bachelor's, postgraduate). The inputs are American Community Survey (ACS) five-year estimates: the 2015--2019 release on 116th-Congress district lines for the 2020 analysis, and the 2020--2024 release on 119th-Congress lines for 2024 (published in January 2026, after the election). Each cell count is the district's total population times the product of the four marginal proportions (Tables B01001, B03002, B15003), so the variables are treated as independent within a district; respondents of other races are in the model as a fourth category but not in the frame. Because the CES filter conditions on turnout, cell counts are converted to expected voters by a separate turnout model (Section~\ref{sec:mrp}).

\subsection{Election results and the lean covariate}
The validation target is the actual Republican share of the two-party presidential vote in each district, from The Downballot's (formerly Daily Kos Elections) calculations of presidential results by congressional district \citep{nir2025downballot}: the 2020 and 2024 results on the lines used in 2024 \citep{downballot2024lines} for the 2024 analysis, and the 2008--2020 results on the lines used in 2020 \citep{downballot2020lines} for the backtest (except North Carolina; Appendix~\ref{app:repro}), with the allocation methodology documented in \citet{downballotmethod}. The MRP model uses the \emph{previous} presidential election as a covariate: the state-level Democratic two-party share and the district's deviation from its state, both standardized. For 2024 these are the 2020 results recalculated onto 119th-Congress lines; for 2020 they are the 2016 results on 116th-Congress lines. The lean covariate therefore never carries information about the election being estimated.

\subsection{Geography}
Table~\ref{tab:data} summarizes both cycles. The two cycles use different district maps. Seven districts that existed in 2020 (CA-53, IL-18, MI-14, NY-27, OH-16, PA-18, WV-03) do not exist in 2024, and Montana's at-large seat became two (MT-01, MT-02). Six new districts were created by 2024 (CO-08, FL-28, NC-14, OR-06, TX-37, TX-38). Every ACS pull and every join in the pipeline asserts the geography it expects; Appendix~\ref{app:repro} describes these checks and what they missed.

\begin{table}[H]
\centering\small
\caption{Data by cycle.}
\label{tab:data}
\begin{tabular}{@{}>{\raggedright\arraybackslash}p{2.05in}>{\raggedright\arraybackslash}p{2 in}>{\raggedright\arraybackslash}p{2in}@{}}
\toprule
 & 2020 (backtest) & 2024 (application) \\
\midrule
Respondents (validated, major-party, complete) & 26,951 & 22,061 \\
Districts & 436 & 436 \\
Respondents per district: median (min--max) & 62 (19--131) & 50 (12--94) \\
Vote validation & Catalist & TargetSmart \\
District lines & 116th Congress & 119th Congress \\
ACS five-year vintage & 2015--2019 & 2020--2024 \\
Prior-election lean covariate & 2016 results & 2020 results, recalculated to new lines \\
Weight variable & \texttt{commonweight}, validated voters & same \\
\bottomrule
\end{tabular}
\end{table}

\section{Methods}

This section describes the methods in the order they are used. Section~\ref{sec:mrp} gives the MRP and poststratification that produce the uncorrected district estimates. Section~\ref{sec:correction} gives the two corrections and how their parameters are estimated on the earlier election. Section~\ref{sec:validation} gives the validation procedure. Section~\ref{sec:lg} gives a model for variation in differential nonresponse across districts, which we fit both to estimate the selection directly and to compare with the corrected MRP. Section~\ref{sec:stageb} gives two variants that apply the correction within poststratification cells. Parameter estimates are reported in Section~4.

\subsection{Multilevel regression and poststratification}\label{sec:mrp}
The preference model is a Bayesian logistic regression fit with \texttt{brms} \citep{burkner2017brms} and Stan \citep{carpenter2017stan}:
\[
\begin{aligned}
\Pr(\text{Republican}_i = 1) = \logit^{-1}\!\big(&\alpha_{\text{age}[i]} + \alpha_{\text{gender}[i]} + \alpha_{\text{race}[i]} + \alpha_{\text{educ}[i]} + \beta_{\text{region}[i]} \\
&+ \gamma_{\text{race}[i]} L_{s[i]} + \delta_{\text{race}[i]} \Delta_{d[i]} + u_{d[i]}\big),
\end{aligned}
\]
where $L_s$ is the standardized prior-election Democratic two-party share of the state, $\Delta_d$ the standardized deviation of the district from its state, $\gamma_{\text{race}}$ and $\delta_{\text{race}}$ race-specific slopes on each, and $u_d$ a district random intercept. The lean covariates are coded as Democratic shares throughout the pipeline (a legacy of its original framing); because the outcome is the Republican vote, their coefficients are negative, and coding them as Republican shares would give an identical fit with opposite-signed coefficients. Demographic effects use explicit reference levels; census region enters as a fixed effect. A turnout model with the same demographic main effects, region, the lean covariates, and state and district intercepts is fit to all respondents in the frame's cells, with validated turnout as the outcome; its predictions, rescaled within state to total votes cast (Appendix~\ref{app:repro}), convert ACS cell counts into expected voters $N_{\text{voter}}$. Each model runs four chains of 2,000 iterations.

The district estimate $\hat\theta_d$ is the $N_{\text{voter}}$-weighted mean of cell-level posterior predictions. Because the model is fit to respondents, $\hat\theta_d$ estimates the Republican share among \emph{respondents} of the district's demographic composition, not among its voters; the difference is the quantity this paper studies.

\subsection{The correction}\label{sec:correction}
The correction addresses two features of the uncorrected estimate that Section~\ref{sec:raw} documents: its level is too low, and its spread is too narrow, because partial pooling shrinks the district estimates toward the regression surface, so the error is more negative in Republican districts and less negative in Democratic ones. We consider two forms.

\paragraph{Share-affine correction.} The first correction, in the style of \citet{heidemanns2020updated}, demeans and stretches in share space,
\[
\theta' = \frac{\theta - \bar b - \lambda/2}{1-\lambda},
\]
where $\bar b$ is the mean bias in the earlier election and $\lambda$ is chosen from a grid so that, in the earlier election, the slope of bias on the true share is close to zero. The stretch is uniform in share space and anchored at 50\%, so the mean residual in the earlier election need not be zero.

\paragraph{Logit-affine correction.} Shrinkage toward a regression surface is a log-odds phenomenon, and a uniform stretch in share space over-corrects the tails while under-correcting the middle. The second correction is affine on the logit scale,
\[
\logit\theta' = \frac{\logit\theta - b}{1-\lambda},
\]
with $(b,\lambda)$ chosen on 2020 so that the regression of bias on the true share and its square has zero linear \emph{and} zero quadratic coefficient. The two conditions identify the two parameters exactly, and we solve them numerically. Under the model of Section~\ref{sec:lg}, $b$ is the logit of the relative response rate for selection conditional on the model's cells, and $\lambda$ corrects the shrinkage, which is a separate phenomenon. Section~\ref{sec:transfer} reports the estimates.

\subsection{Validation}\label{sec:validation}
Validation is a posterior predictive check on the \emph{corrected} quantity. Each of 500 posterior draws (a fixed subsample, seeded) is poststratified and pushed through the correction; the 2.5\% and 97.5\% quantiles form a 95\% posterior predictive band (referred to below as the 95\% band) for each district, and the posterior mean is the point estimate. We report coverage, the number of districts whose actual result lies above versus below the band, the root mean squared error of the point estimate, and the quadratic coefficient of a regression of residual (actual minus estimate) on the estimate and its square. The miss \emph{direction} is the sensitive statistic: a correction with the right level but the wrong shape produces one-sided misses.

The bands include MRP posterior uncertainty but not uncertainty in the transferred parameters, which are treated as fixed; the variants of Section~\ref{sec:stageb} add the latter. As a joint check we use the lineup protocol of \citet{buja2009statistical} and \citet{wickham2010graphical}: each row of the draw matrix is a complete simulated election, and the real election is placed at random among eight simulated ones.

\subsection{A model for variation in differential nonresponse across districts}\label{sec:lg}
\citet{little1998modeling} model differential nonresponse in state polls with a relative response rate $R_j$, the response rate of one candidate's supporters divided by the sum of the two candidates' response rates. We fit this model with districts as the groups. Let $\pi_j$ be the true Republican two-party share, $m_j$ the number of respondents, and $y_j$ the number of Republican respondents. With the number of sampled individuals unknown, $y_j \sim \mathrm{Bin}(m_j, \zeta_j)$ with
\[
\logit\zeta_j = \logit\pi_j + \logit R_j + \logit R^W_j ,
\]
where $R^W_j$ is the part of selection the design weights already encode, estimated from the mean weights of Republican and Democratic respondents in the district (equation 11 of Little and Gelman), with its sampling variance from the delta method, and smoothed across districts by a hierarchical normal model on the logit scale (equation 14 of Little and Gelman), fit by the grid method of \citet{rubin1981estimation}. Districts in which one party has at most one respondent contribute no information to the smoothing and inherit the grand mean.

The hyperparameters $(a_2,b_2)$ of $R_j \sim \mathrm{Beta}(a_2,b_2)$ are estimated on 2020 by maximizing the marginal likelihood with $\pi_j$ fixed at the actual result (equation 13 of Little and Gelman), integrating over $R_j$ on a fixed grid. We estimate them twice: with $R^W_j \equiv 0.5$, giving the \emph{composite} selection, and with the smoothed $R^W_j$, giving the \emph{residual} selection after weighting. The 2024 model is then fit in Stan with $\pi_j \sim \mathrm{Beta}(a_1,b_1)$ under weak hyperpriors on the mean and concentration, $R_j \sim \mathrm{Beta}(\hat a_2,\hat b_2)$ with the 2020 values fixed, and the 2024 $R^W_j$ as data. Four variants are compared: $R \equiv 0.5$ (missing at random), the transferred prior, and the transferred prior with its concentration deflated by one half and one quarter. Calibration is summarized by the statistics Little and Gelman use: the scaled mean of the district $z$-scores and their sum of squares against a $\chi^2_{436}$ reference, computed from posterior $p$-values $\Pr(\pi_j \le \text{actual}_j)$.

\subsection{Applying the correction within poststratification cells}\label{sec:stageb}
The correction of Section~\ref{sec:correction} operates on the poststratified district estimate. The model of Section~\ref{sec:lg} suggests applying it one level down: to each poststratification cell, before aggregation. Per posterior draw, every cell's predicted respondent preference $\zeta_c$ is shifted, $\logit\pi_c = \logit\zeta_c - b_{d(c)}$; cells are poststratified with the same $N_{\text{voter}}$ weights; and the district logits are stretched by $1/(1-\lambda)$ around their weighted mean. Variant 1 uses a constant shift $b$ (two parameters, the same two moment conditions as before). Variant 2 lets the shift vary with Black share centered at its 2020 mean, $b_d = b_0 + b_1(\text{BlackShare}_d - 0.123)$, and adds a third condition: a zero coefficient on Black share in the same regression. Correction-parameter uncertainty enters the 2024 bands through a district bootstrap of the 2020 derivation (500 replicates), one parameter set per posterior draw.

Two design decisions were fixed in advance. The shift $b$ represents selection conditional on the model's demographic cells---the object appropriate to an unweighted MRP---rather than the residual relative response rate after weighting of Section~\ref{sec:lg}, which belongs to a design-weighted variant of the estimator; we treat this as a design choice and note the alternative. And Variant 2's covariate was chosen using an earlier version of the 2024 frame, built on 118th-Congress lines, which differ from the 2024 lines in five states. On that frame, Black share was the one district characteristic whose association with the 2024 residual survived conditioning on the estimate; on the 119th-Congress lines used here, the association is weaker and no longer statistically significant (Section~\ref{sec:survives}). Since the choice was informed by the 2024 residuals, we specified in advance a check---whether 2020 shows the same conditional association---without which Variant 2 cannot transfer. Promotion criteria were also fixed in advance: Variant 1 replaces the district-level correction only if calibration improves at no cost in RMSE; Variant 2 only if it passes the check and removes the association out of sample.

\section{Results}

\subsection{The uncorrected estimate}\label{sec:raw}

\begin{figure}[H]
\centering
\includegraphics[width=0.52\textwidth]{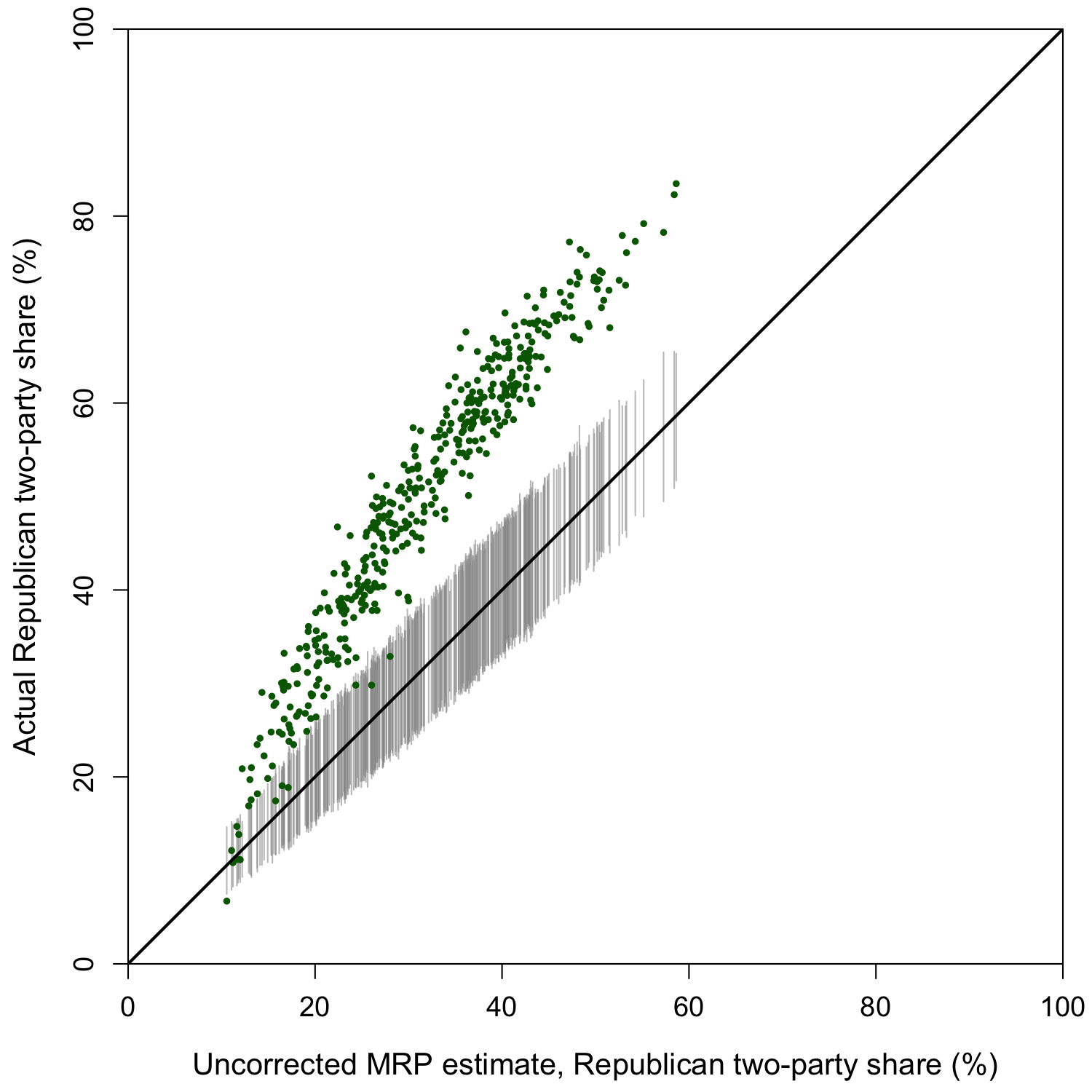}
\caption{Uncorrected 2024 MRP estimate (horizontal) against the actual Republican two-party share (vertical), with 95\% posterior predictive bands. Almost every district lies above its band.}
\label{fig:raw}
\end{figure}

Figure~\ref{fig:raw} below plots the actual Republican two-party vote share in the 2024 election against its uncorrected MRP estimate. It is clear that the MRP estimates are too low in almost every district, with a vast majority of the points lying above the 95\% bands. The only exceptions come in some heavily Democratic districts. A regression of the resulting bias variable (estimate minus actual) on the actual Republican two-party vote share yielded an intercept of $-0.5$ and a slope of $-0.352$ with a $t$-statistic of $-42$ (whose $p$-value is infinitesimal). The MRP estimate is a median of 19 points lower than the actual Republican vote share, and this difference is greater the more Republican a district gets. The 2020 backtest showed the same patterns with a mean bias of $-16.07$ points (i.e. the estimate was on average 16.07 points below the actual 2020 Republican two-party vote share).

\subsection{Transfer of the correction}\label{sec:transfer}

The parameter estimates from the 2020 backtest in the share-affine case were $\bar b = -16.07$ percentage points and $\lambda = 0.40$. For the logit-affine case, they were $b = -0.752$ and $\lambda = 0.275$. In-sample RMSE for the 2020 backtest was 5.58 for share-affine and 4.59 for logit-affine.

\begin{table}[H]
\centering\small
\caption{Out-of-sample validation on 2024, same posterior draws. Misses are districts whose actual result falls outside the 95\% band. The residual quadratic is the coefficient on the squared estimate in a regression of residual on the estimate and its square.}
\label{tab:ab}
\begin{tabular}{lcc}
\toprule
 & Share-affine & Logit-affine \\
\midrule
Coverage (nominal 95\%) & 91.5\% & 98.2\% \\
Misses: actual above band -- below band & 37 -- 0 & 7 -- 1 \\
Residual quadratic & $-0.0058$ & $-0.0028$ \\
RMSE of point estimate (pp) & 5.80 & 4.09 \\
In-sample 2020 RMSE (pp) & 5.58 & 4.59 \\
\bottomrule
\end{tabular}
\end{table}

\begin{figure}[H]
\centering
\includegraphics[width=0.49\textwidth]{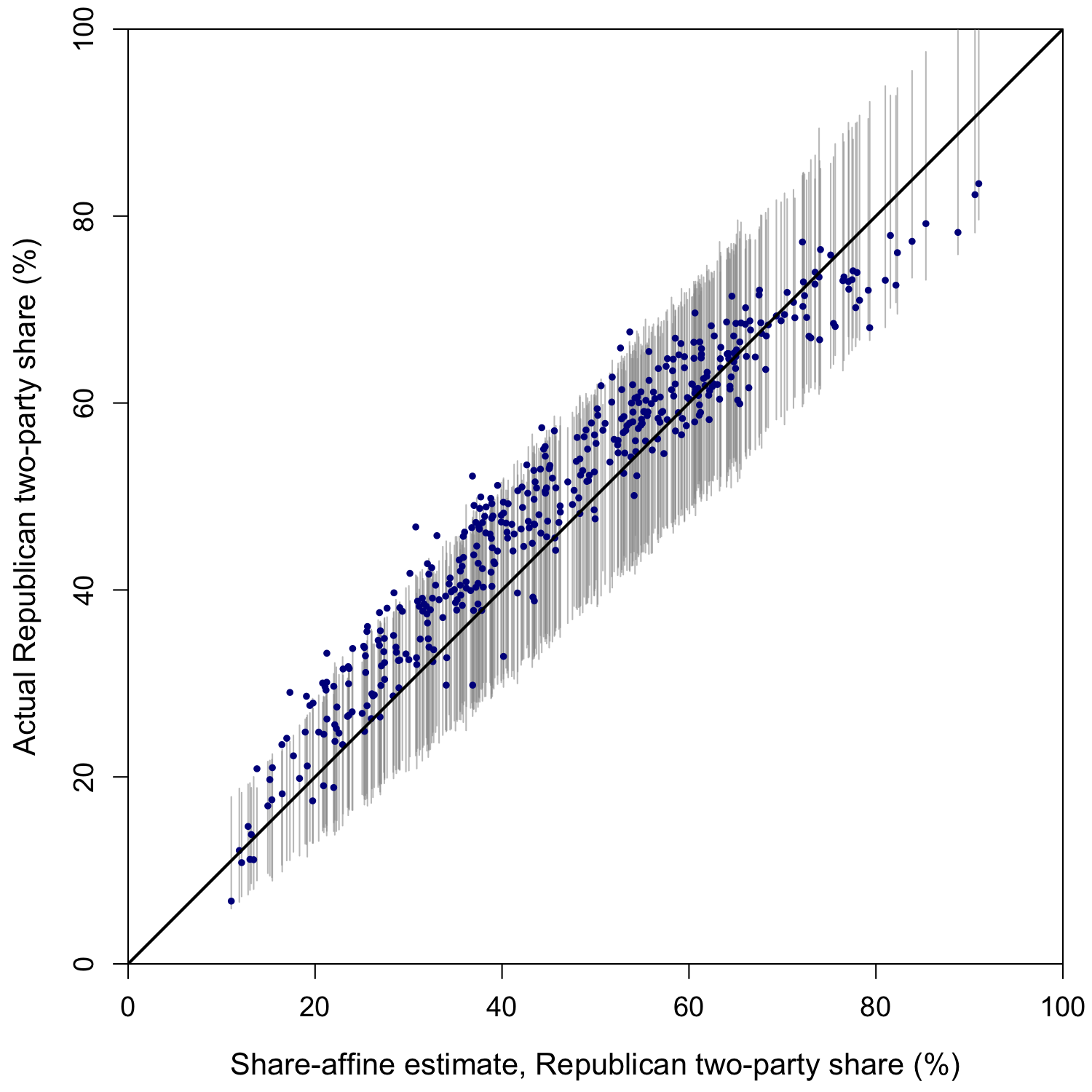}\hfill
\includegraphics[width=0.49\textwidth]{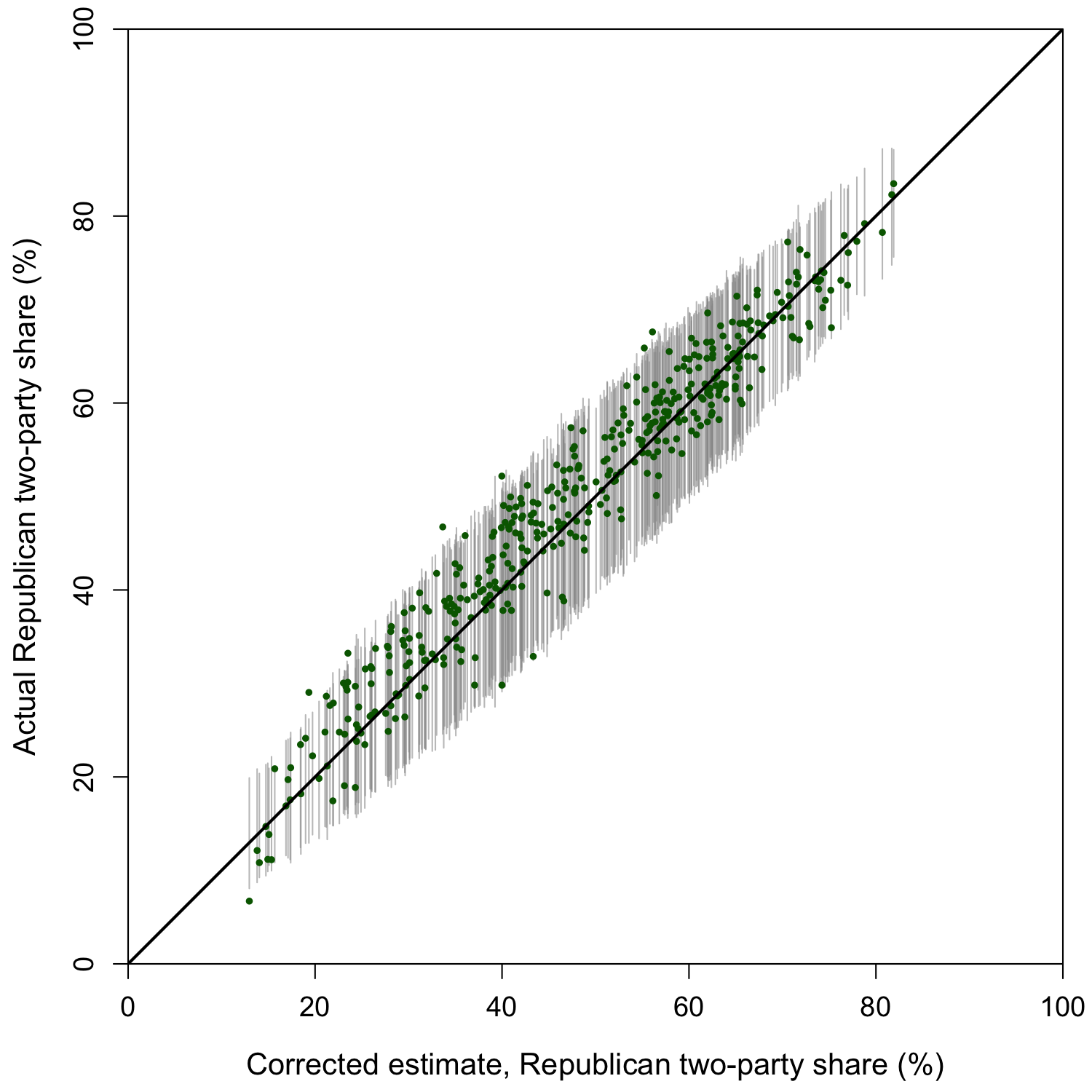}
\caption{Corrected 2024 estimates against the actual result with 95\% posterior predictive bands. Left: share-affine correction (91.5\% coverage; 37 misses, all above the band). Right: logit-affine correction (98.2\%; 7 above, one below).}
\label{fig:ppc}
\end{figure}

We present summary statistics for the two corrections applied to the 2024 election in Table~\ref{tab:ab}. Both corrections use the same seeded posterior draws, so the choice of correction is the only difference between them. 91.5\% of congressional districts lie inside a 95\% band in the share-affine case. Of the 37 districts outside their band, all lie above the band, and they are concentrated in the lowest tercile of the estimate (i.e. the tercile of most Democratic-leaning districts). This estimate had a residual quadratic of $-0.0058$ and an RMSE of 5.80. The logit-affine correction placed 98.2\% of congressional districts inside a 95\% band, and 7 of the 8 district misses lie above that band. The residual quadratic for this method was $-0.0028$ and the RMSE was 4.09. When separated into terciles by estimated Republican vote share, the miss rate by tercile of the estimate was 4.1\% for the lowest tercile, 1.4\% for the middle tercile, and 0\% for the highest tercile. When separated instead into terciles by the actual Republican vote share in the 2024 election, these miss rates are 2.1\%, 2.1\%, and 1.4\%, respectively. The plots of the corrected estimates and their 95\% bands are shown in Figure~\ref{fig:ppc}.

\begin{figure}[H]
\centering
\includegraphics[width=0.57\textwidth]{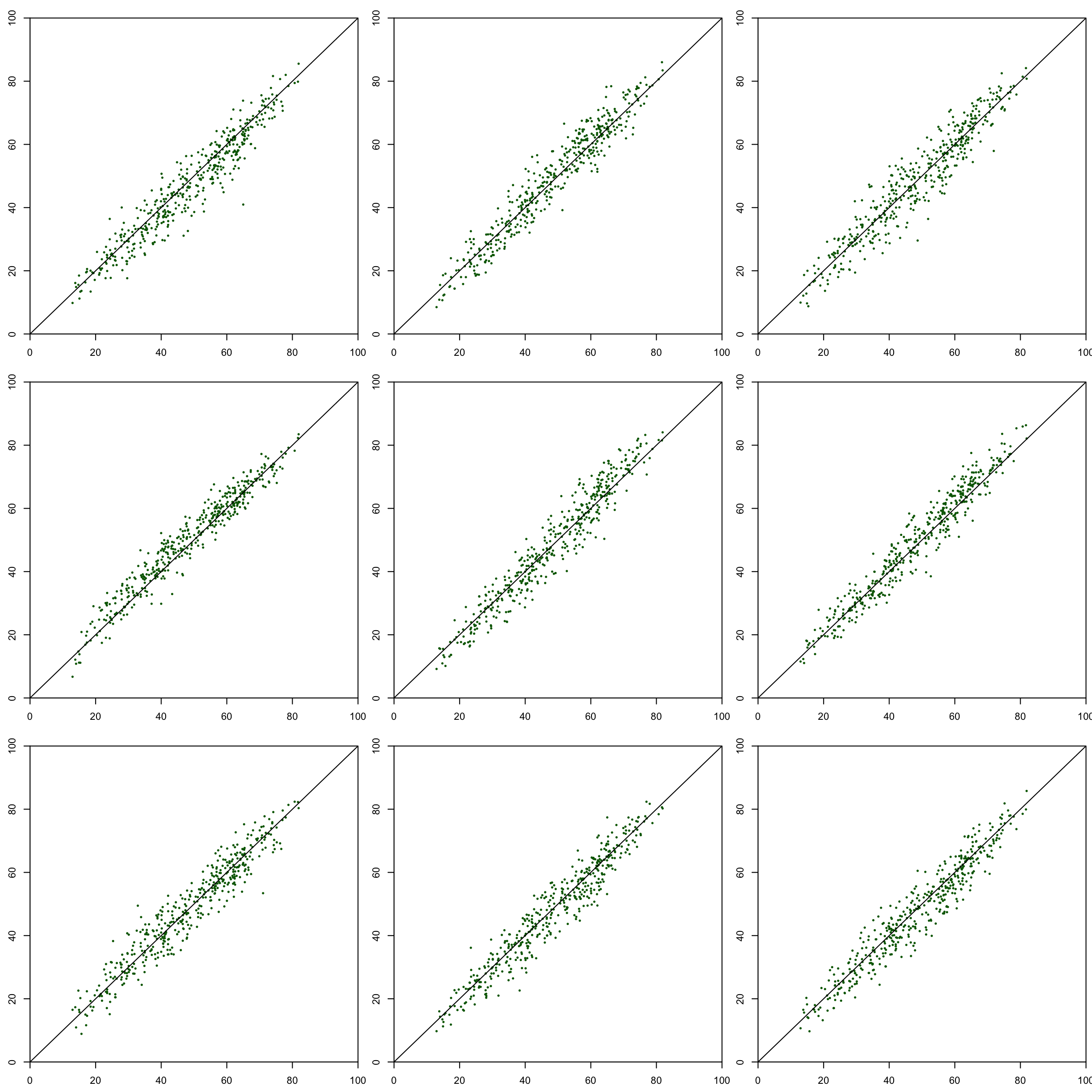}
\caption{Lineup check. Each panel plots a complete election against the logit-corrected estimate; eight are rows of the posterior draw matrix and one is the actual 2024 result (panel 4 in row-major order).}
\label{fig:lineup}
\end{figure}

Figure~\ref{fig:lineup} shows the logit-affine corrected estimate alongside eight simulated posterior draws. The true election result is in panel 4 (in row-major order), and is much less distinguishable than it was before the correction was applied.

Across three MCMC realizations on the earlier frame (Section~\ref{sec:stageb}; two before seeding and one seeded), the number of districts outside the band was 14, 12, and 17, with coverage between 96.1\% and 97.2\%; the direction, curvature, and RMSE all had negligible variance across the draws. We report the counts from the seeded run made after the corrections described in Appendix~\ref{app:repro} as the ``official'' results. The asymmetry between overestimates and underestimates held in every realization (12--2, 10--2, and 16--1 above and below the band); for the official run, 7 above and 1 below gives $p \approx 0.07$ under a null hypothesis of symmetry, so the count alone is weak evidence; the mean residual in Section~\ref{sec:survives} is the stronger sign that the estimates run low.

\subsection{What remains after the correction}\label{sec:survives}

Table~\ref{tab:curve} shows the decomposition of the residual curvature. All three columns are measured against the share-affine estimate and its square, so they can be compared directly, and the logit-affine residual is exactly the share-affine residual minus what the logit-affine correction removes. Because the share-affine correction is linear in the uncorrected estimate, ratios of these coefficients do not depend on its parameters and describe the uncorrected estimate. 

\begin{table}[H]
\centering\small
\caption{Residual curvature on a common axis. Entries are coefficients on the squared estimate from regressions on the share-affine estimate and its square: of the share-affine residual, of the logit-affine residual, and of the difference between the logit-affine and share-affine estimates (the curvature that the logit-affine correction removes). The third column is the first minus the second, before rounding.}
\label{tab:curve}
\begin{tabular}{@{}>{\raggedright\arraybackslash}p{1.9in}>{\centering\arraybackslash}p{1.3in}>{\centering\arraybackslash}p{1.5in}>{\centering\arraybackslash}p{1.3in}@{}}
\toprule
 & Share-affine residual & Logit-affine residual & Removed by logit-affine correction \\
\midrule
2020 (in-sample) & $-0.00380$ & $+0.00005$ & $-0.00385$ \\
2024 (out-of-sample) & $-0.00584$ & $-0.00179$ & $-0.00405$ \\
\midrule
Share of 2024 curvature & 100\% & 31\% & 69\% \\
\bottomrule
\end{tabular}
\end{table}

For the 2020 in-sample cycle, although $\lambda$ was chosen on the same 2020 data to bring the linear term close to zero (a slope of $+0.066$ remains), regressing the in-sample residual of the share-affine correction on the estimate and its square gives a quadratic coefficient of $-0.0038$ ($t = -6.9$). The fitted residual peaks at $+4.3$ points near an estimate of 28\%. The logit-affine correction, whose two parameters were chosen on 2020, removes all of this curvature in-sample. In 2024, the share-affine residual has a quadratic coefficient of $-0.0058$, roughly 1.5 times the 2020 value, so 65\% of the 2024 curvature was already present in 2020. Applied out-of-sample, the logit-affine correction removes 69\% of the 2024 curvature and leaves 31\%, which is specific to 2024. Table~\ref{tab:ab} reports each correction's residual curvature against its own estimate ($-0.0058$ and $-0.0028$); those two coefficients are on different axes and do not enter this decomposition.

The mean residual under the logit correction was $+1.62$ points (with a $p$-value below 0.001), compared with $-0.07$ points in-sample in 2020; under the share-affine correction, it was $+3.42$ points, compared with $+1.82$ points in-sample, an offset that comes from anchoring the stretch at 50\% (Section~\ref{sec:correction}). Under either correction the mean residual rose by 1.6 to 1.7 points between the two elections; this is the level shift.

\begin{table}[H]
\centering\small
\caption{Regressions of the 2024 residual (actual minus logit-corrected estimate, pp) on district composition. Standard errors in parentheses.}
\label{tab:resid}
\begin{tabular}{lcc}
\toprule
 & Marginal & Conditional on estimate \\
\midrule
Intercept & $1.22$ (0.33) & $-2.01$ (1.61) \\
Estimate & -- & $0.261$ (0.066) \\
Estimate$^2$ & -- & $-0.00340$ (0.00068) \\
Hispanic share & $0.04$ (1.02) & $-2.25$ (1.03) \\
Black share & $3.27$ (1.43) & $2.04$ (1.57) \\
$R^2$ & 0.012 & 0.117 \\
\bottomrule
\end{tabular}
\end{table}

Regressions of the 2024 residual on district Hispanic and Black population shares are shown in Table~\ref{tab:resid}. Without conditioning on the estimate and its square, Black share has a marginal coefficient of $+3.27$ ($p = 0.023$) and Hispanic share has a marginal coefficient of $+0.04$ ($p = 0.97$), both in residual points per unit of population share. Under the share-affine correction, the Hispanic coefficient had been $+3.0$ and significant; therefore, the logit correction effectively accounted for this variability. Conditional on the estimate and its square, the Black coefficient was $+2.04$ ($p = 0.19$) and the Hispanic coefficient was $-2.25$ ($p = 0.029$). 

Table~\ref{tab:misses} in Appendix~\ref{app:tables} lists the districts that lie outside the 95\% band after the logit-affine correction, and Figure~\ref{fig:labeled} plots every district with its label. For districts above the band, the median two-party swing from 2020 to 2024 was 4.1 percentage points, compared to 2.2 percentage points for districts inside the 95\% band. The upper quartile of districts above the 95\% band was 8.2 percentage points. Three of these districts (NY-06, FL-26, and NY-03) are in the top tenth of all districts by 2020-to-2024 Republican swing, while two others (AL-02 and KS-01) swung by less than the median district. The one district below its 95\% band is DC-01, where the predicted two-party vote share for Republicans was 13.0\%; in reality, it was 6.7\%. Districts outside the band have a median of 35.5 respondents in the CES survey, compared with 50 for districts inside the band, but three of the eight have more than 55. Therefore, we cannot say that small samples alone explain the missed district estimates.

\subsection{Variation in differential nonresponse across districts}\label{sec:lgresults}

\begin{figure}[H]
\centering
\includegraphics[width=0.88\textwidth]{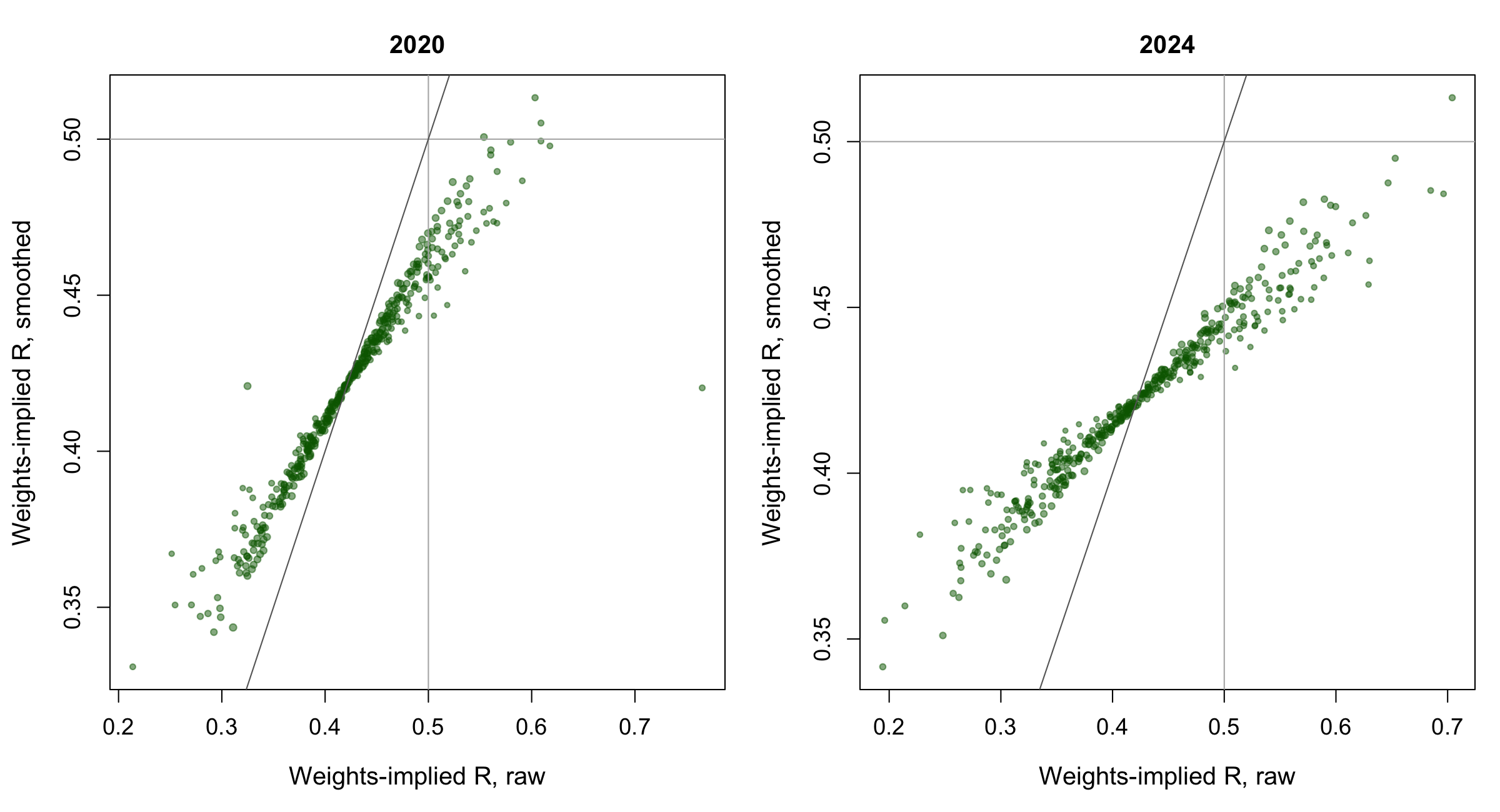}
\caption{Weights-implied selection $R^W_j$ by district, raw (horizontal) against hierarchically smoothed (vertical), for 2020 and 2024. Point size is proportional to $\sqrt{m_j}$. Nearly every district lies below 0.5: the design weights correct for Republican under-selection throughout.}
\label{fig:rw}
\end{figure}

Figure~\ref{fig:rw} plots the weights-implied adjustment $R_j^W$ before smoothing (x-axis) against its smoothed values (y-axis). $R_j^W$ has a before-smoothing mean of 0.425 in both 2020 and 2024. After smoothing, it has a grand mean of $R_{\mathrm{adj}} = 0.421$ in 2020 and $R_{\mathrm{adj}} = 0.422$ in 2024, with a between-district standard deviation $\tau \approx 0.20$ on the logit scale. This shows that the CES data in both elections weigh Republican respondents 1.375 times as much as Democratic respondents $\big(\frac{1-0.421}{0.421} \approx 1.375\big)$. With $\pi_j$ fixed at the 2020 results, the composite selection parameter estimate, from the counts with the weights ignored, is $R_{\mathrm{comp}} = 0.375$ (Beta(53.8, 89.6), with the standard deviation across districts being 0.040). This means that in a tied district, unweighted respondents would be 37.5\% Republican; equivalently, Republican voters enter the sample at $e^{\operatorname{logit}(0.375)} = e^{-0.511} \approx 0.60$ times the odds of Democratic voters. Estimating from the weights alone without using the results, we have from before that $R_{\mathrm{adj}} = 0.421$: applying the weights multiplies the Republican odds by 1.375. Estimating from the counts given the weights, the residual selection parameter estimate is $R_{\mathrm{res}} = 0.454$ (Beta(110.1, 132.6), with the standard deviation across districts being 0.032). Therefore, in a tied district, weighted respondents would still be just 45.4\% Republican. If DC-01 is excluded, the residual estimate changes slightly to Beta(110.3, 132.8).

Of the three $R$ values we reported, two of them (the composite selection parameter and the weights-implied adjustment) use exactly one predictor: the vote counts for the composite selection parameter and the weight ratios for the weights-implied adjustment. The third, the residual selection parameter, uses the counts given the weights. Under the model of Section~\ref{sec:lg}, it should be the case that $R_{\mathrm{comp}}$ is the sum of the other two on the log-odds scale: $\operatorname{logit}(R_{\mathrm{comp}}) = \operatorname{logit}(R_{\mathrm{adj}}) + \operatorname{logit}(R_{\mathrm{res}})$. In this case, $\operatorname{logit}(R_{\mathrm{adj}}) + \operatorname{logit}(R_{\mathrm{res}}) = \operatorname{logit}(0.421) + \operatorname{logit}(0.454) \approx -0.319 - 0.185 = -0.504$ and $\operatorname{logit}(R_{\mathrm{comp}}) = \operatorname{logit}(0.375) \approx -0.511$. Therefore, the composite selection estimated directly from the counts is almost entirely reproduced by the weight ratios and the residual together; the small difference is estimation error, not a component of the selection. We then have that Republican voters enter the sample at 0.60 times the odds of Democratic voters, and these weights account for $\frac{\operatorname{logit}(0.421)}{\operatorname{logit}(0.375)} = \frac{-0.319}{-0.511} \approx 62\%$ of the selection on the log-odds scale. The remainder is attributed to residual differential nonresponse. 

For comparison, \citet{little1998modeling} found $R_\mathrm{comp} = 0.534$ and $R_\mathrm{res} = 0.515$ in a similar study on CBS telephone polls prior to the 1988 presidential election; in this case, the opposite effect was observed, where supporters of the Republican candidate (George H.W. Bush) would have made up 53.4\% of respondents in an evenly divided state. The standard deviation across states for the 1988 study was 0.044, close to the 0.040 we observed across districts for the 2020 CES. We then know that the weights account for $\frac{\operatorname{logit}(R_\mathrm{adj})}{\operatorname{logit}(R_\mathrm{comp})} = \frac{\operatorname{logit}(R_\mathrm{comp}) - \operatorname{logit}(R_\mathrm{res})}{\operatorname{logit}(R_\mathrm{comp})} = \frac{\operatorname{logit}(0.534) - \operatorname{logit}(0.515)}{\operatorname{logit}(0.534)} \approx \frac{0.076}{0.136} \approx 56\%$ of the selection on the log-odds scale (as reported in their text), a figure similar to our 62\% from 2020.

\begin{table}[H]
\centering\small
\caption{Eight estimators of the 2024 district Republican share, evaluated the same way. RMSE is on the proportion scale and uses posterior medians as point estimates, so the MRP rows differ slightly from Tables~\ref{tab:ab} and~\ref{tab:stageb}, which use posterior means in percentage points. Mean $z$ is the scaled mean of district $z$-scores computed from posterior $p$-values; $\chi^2$ is their sum of squares, against 436 degrees of freedom. The two variant rows are the only MRP-based rows whose bands include uncertainty in the correction parameters.}
\label{tab:lg}
\begin{tabular}{lcccccc}
\toprule
Model & RMSE & Coverage & Above -- below & Mean $z$ & $\chi^2$ & $p$ \\
\midrule
Raw CES (Jeffreys) & 0.134 & 63.1\% & 160 -- 1 & $+32.7$ & 1646 & $10^{-140}$ \\
Nonresponse model, $R \equiv 0.5$ & 0.085 & 83.9\% & 57 -- 13 & $+11.9$ & 837 & $10^{-27}$ \\
Nonresponse model, 2020 prior & 0.078 & 90.1\% & 16 -- 27 & $-2.3$ & 609 & $8\times10^{-8}$ \\
Nonresponse model, prior $\times\tfrac12$ conc. & 0.080 & 91.3\% & 15 -- 23 & $-2.3$ & 579 & $5\times10^{-6}$ \\
Nonresponse model, prior $\times\tfrac14$ conc. & 0.083 & 93.3\% & 10 -- 19 & $-2.2$ & 555 & $10^{-4}$ \\
MRP + logit correction & 0.041 & 98.2\% & 7 -- 1 & $+6.9$ & 338 & 1.00 \\
Variant 1 (cell-level shift) & 0.044 & 96.8\% & 13 -- 1 & $+8.2$ & 348 & 1.00 \\
Variant 2 (+ Black-share term) & 0.046 & 97.5\% & 0 -- 11 & $-9.8$ & 358 & 1.00 \\
\bottomrule
\end{tabular}
\end{table}

\begin{figure}[H]
\centering
\includegraphics[width=0.88\textwidth]{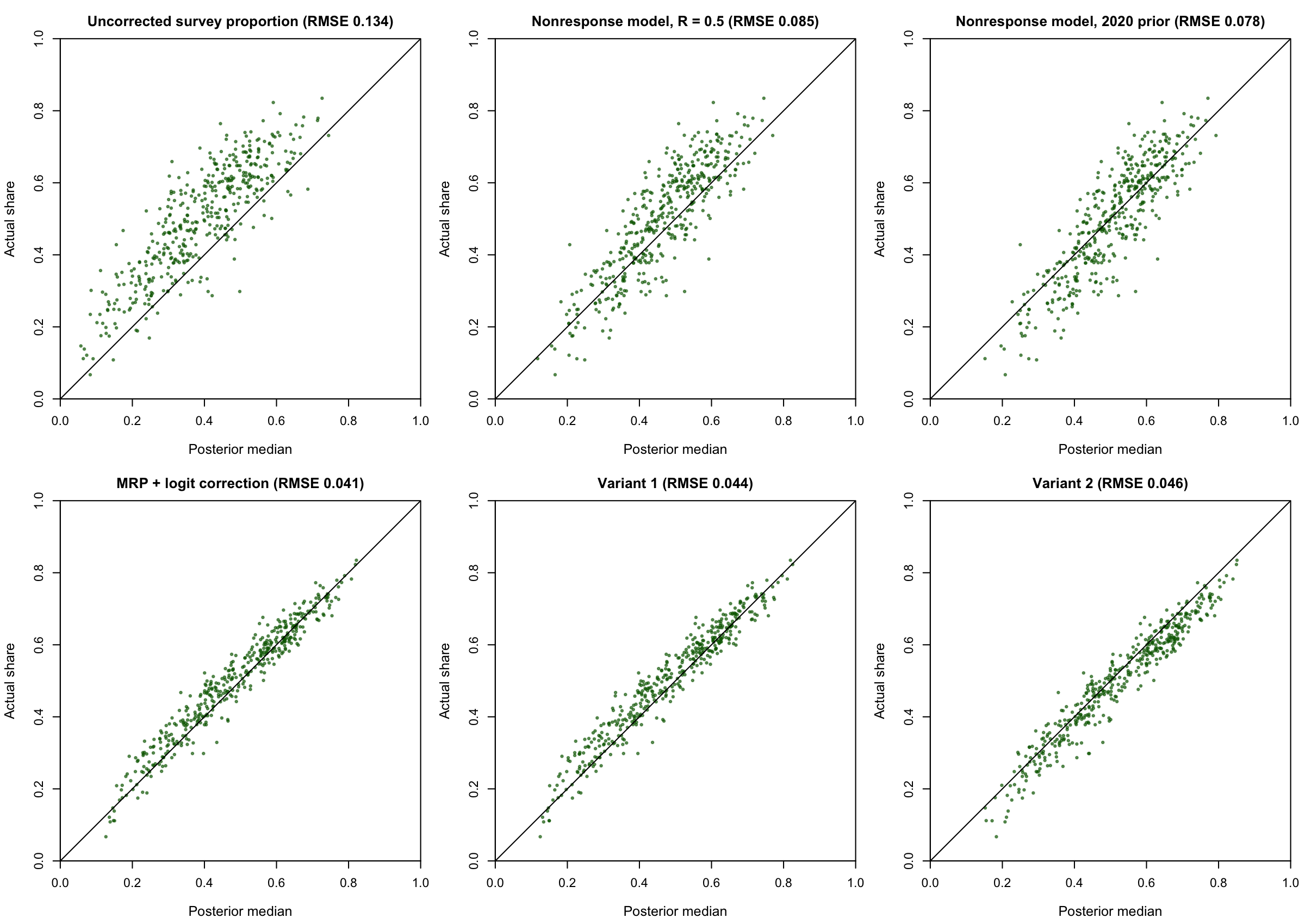}
\caption{Actual 2024 Republican share (vertical) against the posterior median (horizontal) for six estimators: the uncorrected survey proportion, the nonresponse model with $R$ fixed at 0.5, the nonresponse model with the 2020 prior, MRP with the logit correction, and the two cell-level variants.}
\label{fig:strip}
\end{figure}

Table~\ref{tab:lg} compares the performance of eight different estimators of 2024 district Republican vote share. We evaluate each based on how closely the actual results resemble what the model's posteriors say they should be: a 95\% coverage close to 95\%, roughly equal misses above and below, a mean $z$ that is close to 0 (distributed $\mathrm{N}(0,1)$), and a $\chi^2$ close to the number of congressional districts (distributed approximately $\mathrm{N}(436,\,30^2)$). The $R \equiv 0.5$ nonresponse model, which assumes that nonresponse bias is ignorable, fails each of these tests: not enough estimates are within the 95\% band, too many of the misses are above the predicted range, the mean $z$ is too high (i.e. the actual results are on average much greater than the estimates), and the $\chi^2$ statistic is too far from 436, indicating there is not enough spread for this model to be reasonable. Adding a 2020 prior to the nonresponse model brings the mean $z$ to a reasonable distance from 0, but does not reduce the $\chi^2$ statistic by enough to fix the spread issue and also does not sufficiently improve the coverage percentage. These results are almost identical when the 2020 prior's concentration is halved or quartered, indicating that the nonresponse model's limitation is instead the exchangeable Beta prior on $\pi_j$, which cannot use district demographics and over-shrinks the tails (as shown in Figure~\ref{fig:strip}; Figure~\ref{fig:errm} shows that its error does not narrow with district sample size). Our MRP with the logit correction improves the RMSE to 0.041 (compared to 0.078 in the best nonresponse model fit), has a higher coverage percentage (98.2\%) than any of the nonresponse models, and the $\chi^2$ statistic improves to about three standard deviations from the theoretical number (436). However, the mean $z$ is much higher than we would expect it to be and outside a reasonable distance from 0. Therefore, our MRP+logit model suffers from the opposite issue to the nonresponse model: its bands have about the right width, but its level is off: the estimates are 1.62 points too low on average (Section~\ref{sec:survives}), because the shift estimated on 2020 is too small for 2024. In Section~\ref{sec:stagebresults}, we try to improve on these results with two variant models that combine the two approaches.

\subsection{Cell-level variants}\label{sec:stagebresults}

The correction in Section~\ref{sec:correction} is applied to the
district-level MRP estimate, after poststratification. However,
differential nonresponse occurs at the individual level, suggesting
that the correction might instead be applied within poststratification
cells, before they are aggregated into district estimates.
Section~\ref{sec:stageb} describes two such variants. Variant 1 applies
a common shift $b$ to every cell's predicted preference on the
log-odds scale, followed by the usual district-level stretch. Variant
2 allows the shift to vary with the district's Black population share,
$b_d=b_0+b_1(\text{BlackShare}_d-0.123)$, to accommodate the possibility
that the selection mechanism differs with district racial composition.

The distinction between the district-level and cell-level approaches
is consequential because correction and poststratification do not
generally commute. Let $p_{dc}$ denote the predicted Republican
preference in cell $c$ of district $d$, and let $w_{dc}$ denote its
poststratification weight. The uncorrected district estimate is
\[
\hat p_d=\sum_c w_{dc}p_{dc}.
\]
The district-level correction first aggregates the cell predictions
and then applies the logit-affine transformation. The cell-level
variants instead shift the cell predictions before aggregation and
then apply the district-level stretch. Because the logit transformation
is nonlinear, these procedures need not produce the same corrected
district estimate, even when they use the same shift parameter.

Table~\ref{tab:stageb} reports the parameter estimates obtained using
the 2020 data and the corresponding validation results on 2024. Both
variants propagate uncertainty in the correction parameters through
a bootstrap of 500 district resamples from 2020, with a separate
parameter set used for each 2024 posterior draw.

\begin{table}[H]
\centering\small
\caption{Cell-level variants: parameters estimated on 2020, and validation on 2024 with correction-parameter uncertainty from a 500-replicate district bootstrap. The conditional Black-share coefficient is from the regression of each column's residual (actual minus estimate) on its estimate, the square of the estimate, and Hispanic and Black shares. Standard errors in parentheses.}
\label{tab:stageb}
\begin{tabular}{@{}>{\raggedright\arraybackslash}p{2.1in}>{\centering\arraybackslash}p{1.3in}>{\centering\arraybackslash}p{1.3in}>{\centering\arraybackslash}p{1.3in}@{}}
\toprule
 & District-level correction & Variant 1 (cell-level) & Variant 2 (+ Black share) \\
\midrule
$b$ (or $b_0$) & $-0.752$ & $-0.796$ (0.081) & $-1.023$ \\
$b_1$ per unit Black share & -- & -- & $-0.832$ (0.090) \\
$\lambda$ & 0.275 & 0.290 (0.012) & 0.327 \\
2020 in-sample RMSE & 4.59 & 4.84 & 6.01 \\
2020 conditional Black-share coefficient & $-0.1$ & $+6.9$ & $-8.0$ \\
\midrule
2024 coverage & 98.2\% & 96.8\% & 97.5\% \\
2024 misses above -- below & 7 -- 1 & 13 -- 1 & 0 -- 11 \\
2024 RMSE & 4.09 & 4.35 & 4.62 \\
2024 conditional Black-share coefficient & $+2.0$ & $+7.4$ & $-12.0$ \\
\bottomrule
\end{tabular}
\end{table}

Neither variant improves on the district-level correction according
to the criteria specified in advance. Variant 1 would replace the
district-level correction only if it improved calibration without
increasing RMSE. Variant 2 would replace it only if it first passed a
pre-specified check on 2020 and then removed the conditional
association between Black population share and the 2024 residual.
This additional check is particularly important because Black
population share was selected as a covariate after examining the
2024 residuals on the earlier frame. Although the 2024
observations were not used to estimate the correction parameters,
using their residual pattern to choose the model specification creates
a risk of overfitting. We therefore require the corresponding
conditional relationship to be present in 2020 before treating the
proposed correction as transferable to 2024.

Table~\ref{tab:gate} reports the regressions used for this check. The
first column regresses the 2020 in-sample residual of the district-level
correction on the corrected estimate, its square, and both Black and
Hispanic population shares. The second column repeats
this regression after excluding the 44 districts in the top
decile of Hispanic population share (Table~\ref{tab:hisp}), including
20 districts in California and 11 in Texas.

\begin{table}[H]
\centering\small
\caption{Regressions of the 2020 in-sample residual of the logit-affine correction (actual minus estimate, pp) on the estimate, its square, and district composition, used in the diagnosis of Variant 2. Standard errors in parentheses.}
\label{tab:gate}
\begin{tabular}{lcc}
\toprule
 & All districts (436) & Excluding top-decile Hispanic (392) \\
\midrule
Intercept & $4.56$ (1.58) & $2.24$ (1.62) \\
Estimate & $-0.003$ (0.063) & $0.040$ (0.063) \\
Estimate$^2$ & $-0.0011$ (0.0007) & $-0.0014$ (0.0007) \\
Hispanic share & $-9.87$ (1.16) & $-2.85$ (2.13) \\
Black share & $-0.11$ (1.74) & $1.28$ (1.69) \\
\bottomrule
\end{tabular}
\end{table}

The results reveal three features of the residual structure that help
explain the performance of the cell-level variants.

First, the conditional association between Black population share
and the 2024 residual that motivated Variant 2 is not reproduced in 2020. In the
first column of Table~\ref{tab:gate}, the Black-share coefficient is
$-0.11$ (standard error 1.74; $p=0.95$), after controlling for
Hispanic share and the corrected estimate. The pre-specified check
for Variant 2 is therefore not satisfied. In 2024, the corresponding
coefficient is $+2.0$ (standard error 1.57; $p=0.19$; Table~\ref{tab:resid}); on the
earlier frame from which the covariate was chosen (Section~\ref{sec:stageb}), it was $+3.2$
($p=0.041$). The 2020 and 2024 coefficients differ in sign, but neither is
statistically distinguishable from zero. This comparison does not, by itself, establish a
change in the voting behavior of individual racial groups, nor does
it constitute a formal test of the difference between the two
coefficients. It does show that the relationship that motivated Variant 2 is not evident
in the 2020 data from which the correction is estimated.

Second, the 2020 residuals exhibit a substantially stronger
association with Hispanic population share. In the first column of
Table~\ref{tab:gate}, the Hispanic-share coefficient is $-9.87$,
with a $t$-statistic of approximately $-8.5$. Since the dependent
variable is actual minus estimated Republican support, this indicates
that the corrected estimate is too Republican in districts with
larger Hispanic populations, conditional on the other variables.
The corresponding coefficient in 2024 is only $-2.25$
(Section~\ref{sec:survives}), less than one-quarter as large.
Moreover, the 2020 association appears particularly strong among
the most Hispanic districts. When the 44 districts in the top decile
of Hispanic population share are excluded, the coefficient falls
from $-9.87$ to $-2.85$ and is no longer statistically significant
($p=0.18$). The reduction may partly reflect the narrower range of
Hispanic share in the remaining districts; it is also consistent
with the original association being disproportionately influenced
by the most Hispanic districts. The regression comparison alone
cannot establish the precise functional form of this relationship,
but it suggests that a single linear adjustment for Hispanic share
may not adequately describe the 2020 residual pattern. In any case,
the much smaller 2024 coefficient indicates that this relationship
does not transfer at the same magnitude across the two elections.

Third, the additional condition of Variant 2 is not written in the
regression used for the check. The check regresses the residual on
the estimate, its square, and Hispanic and Black shares. The
condition sets to zero the Black-share coefficient in the regression
of Section~\ref{sec:stageb}, which conditions on the actual share
and its square and does not include Hispanic share. For Variant 1,
the Black-share coefficient of the 2020 residual is $+6.9$ in the
check regression (Table~\ref{tab:stageb}), an association that the
cell-level shift itself creates, since the district-level correction
has $-0.1$. Omitting Hispanic share raises it to $+11.1$, because
Black share partially absorbs the association with Hispanic share,
and conditioning on the actual share rather than the estimate raises
it to $+15.4$ (standard error 1.8). The condition removes all of
this, with $b_1=-0.832$. In the check regression, the result is a
coefficient of $-8.0$ for Variant 2's own 2020 estimates: the variant
over-adjusts in the election from which it was estimated. This contributes
to Variant 2's deterioration in 2020 in-sample RMSE, from 4.59 to
6.01 percentage points.

The 2024 results provide further evidence that Variant 2
over-adjusts. Its conditional Black-share coefficient changes from
$+2.0$ under the district-level correction to $-12.0$, reversing
the direction of the original association. All eleven districts outside
its 95\% bands fall below the lower bound, with none above the
upper bound. Although Variant 2 has high nominal coverage
(97.5\%), this figure masks a pronounced one-sided pattern of
misses. Its mean standardized $z$-score is $-9.8$
(Table~\ref{tab:lg}), indicating that the actual results tend
to fall below the model's predictions. Coverage alone is therefore
insufficient to establish good calibration: the direction of the
misses, the distribution of standardized residuals, and point-estimate
RMSE also matter. Variant 2 has a 2024 RMSE of 4.62, compared with
4.09 for the district-level correction, and its residual pattern
suggests systematic over-adjustment.

Finally, Variant 1 also performs worse than the district-level
correction, despite using nearly identical estimated parameters.
Its 2020 in-sample RMSE increases from 4.59 to 4.84 percentage
points, while its 2024 RMSE increases from 4.09 to 4.35. The
conditional Black-share coefficient in the 2024 residual also
increases from $+2.0$ to $+7.4$. Thus, applying the correction
within cells does not remove the residual association with district
racial composition; in this specification, the association becomes
more pronounced.

The difference between the two approaches arises from the nonlinear shape of the logit-shift transformation. For a positive shift \(a=-b\), define
\[
g(p)=\operatorname{logit}^{-1}\{\operatorname{logit}(p)+a\}.
\]
When \(a>0\), \(g(p)\) is concave in \(p\). Consequently, for poststratification weights that sum to one,
\[
\sum_c w_{dc}g(p_{dc})
\;\leq\;
g\left(\sum_c w_{dc}p_{dc}\right).
\]
Thus, for a common shift, applying the adjustment within cells before poststratification generally produces a smaller aggregate increase than applying the same shift to the district-level prediction. The difference is more pronounced when cell-level predictions are heterogeneous.

This provides a possible explanation for the 2024 results: the cell-level correction leaves a larger conditional association between Black population share and the residual (\(+7.4\), compared with \(+2.0\) under the district-level correction). Since the residual is defined as actual minus estimated Republican support, the positive coefficient indicates greater underestimation in districts with larger Black population shares. The cell-level variant therefore does not address this pattern as effectively as the district-level correction. The concavity argument explains why the two procedures can produce different adjustments, while the residual regressions provide the empirical evidence that Variant 1 performs worse in this application.

Taken together, these results do not establish that cell-level correction is inherently inappropriate, or that a common log-odds shift could never describe differential selection across cells. Rather, they show that the particular specifications considered here do not improve on the district-level correction. Variant 1 illustrates how the nonlinear interaction between correction and poststratification can affect aggregate adjustments, while Variant 2 illustrates the risks of introducing demographic interactions based on residual relationships that are not stable across elections. More generally, exact moment matching constrains only the regression in which the condition is written; it does not guarantee that the relationship is absent from the regression used to judge the estimate, or that it will persist in a new election. With only two elections available, it is difficult to distinguish persistent features of differential nonresponse from election-specific residual patterns. Neither cell-level variant therefore provides sufficient evidence to replace the simpler district-level correction.

\section{Discussion}

The level and the compression of the uncorrected estimate are stable properties of the survey and of the estimation method. A correction estimated on one presidential election carries them to the next, and the corrected bands have close to nominal coverage. What does not transfer is what changed in the electorate between the two elections: the level of selection increased slightly, and several of the districts in which the Republican vote grew fastest between 2020 and 2024 lie outside the corrected bands. This is the district-level counterpart of a finding from 2016, when departures from a national swing were small in aggregate but decisive where they occurred \citep{gelman2017nineteen}. A panel is least likely to have recruited the newest members of a coalition, so the survey's selection changes as the electorate changes.

The relative response rate of Section~\ref{sec:lg} and the data-defect correlation of \citet{meng2018statistical} measure the same selection in different units. By Meng's identity, the error of a sample mean is the product of the correlation between responding and the outcome, the standard deviation of the outcome, and a factor that depends on the sampling fraction. A composite $R$ of 0.375 implies an under-estimate of about 12.5 points at an evenly divided district, which at the sampling fractions of the CES corresponds to a correlation of about $-0.003$ to $-0.004$. This is the range \citet{chooi2026persistent} reports for 2024 ($-0.0030$ for validated voters, $-0.0044$ for all respondents) and \citet{meng2018statistical} reports for 2016 ($-0.0045$). The correspondence is approximate because their outcome is Trump's share of all respondents and ours is the two-party share among validated voters. Our finding that small samples alone do not explain the districts outside the band is consistent with their result on effective sample size: under differential nonresponse of this magnitude the effective sample is a dozen or two respondents per state regardless of the nominal count, and additional respondents from the same panel add little. Their correction is a constant national shift, which corresponds to our level parameter without the shape parameter. Their transfer across eight years and ours across four together suggest that the level of the bias has been roughly stable across at least three election cycles; our level shift of 1.7 points shows that it is not exactly stable.

Our pipeline poststratifies an unweighted MRP, so the natural selection parameter is $R$ conditional on the model's cells, net of what poststratification absorbs. A design-weighted variant of the estimator would instead use the residual relative response rate after weighting, estimated in Section~\ref{sec:lgresults}. Which is preferable depends on what the CES weights are raked on and how that overlaps with the MRP covariates, a question the survey's designers are better placed to answer than we are. It is also the question on which weighting-based corrections rest, since they require selection to be ignorable given the raking variables \citep{bailey2023new}. We have stated the choice explicitly and built the variants so that the alternative is a single substitution.

Two methodological points follow from the variants. First, Variant 2's covariate was chosen by inspecting the 2024 residuals, which are the test set, as computed on the earlier frame (Section~\ref{sec:stageb}); on the 119th-Congress lines the association that prompted the choice is no longer statistically significant. No numbers from 2024 entered the fit, but the model form did, and the check on 2020 that we specified in advance is what caught it. Second, a moment condition constrains only the regression in which it is written. Variant 2's third condition, like the first two, is written in a regression on the actual share, and it is met exactly; the check conditions on the estimate, and in that regression the fitted variant has a Black-share coefficient of $-8.0$ in 2020 itself (Section~\ref{sec:stagebresults}). Running the check on the fitted variant, and not only on the district-level correction, would have shown the over-adjustment before 2024 was examined.

The study has several limitations. The transfer is tested on one pair of elections, and that pair spans a change in the racial composition of the two coalitions, so the instability of the race-related terms in the residual may be a property of this pair rather than of the survey. The poststratification frame uses total population by race and Hispanic origin rather than citizen voting-age population, which may contribute to the over-estimate of the Republican share in the most Hispanic districts. It also takes age, gender, race, and education to be independent within a district. The 2024 frame comes from an ACS release published after the election; a pre-election application would have to use an earlier release and, in the five states that redrew their districts, build the frame from smaller geographic units. In the 2020 backtest, North Carolina is analyzed on the 116th-Congress lines rather than the map on which its 2020 election was held, with results aggregated from precinct returns (Appendix~\ref{app:repro}). The outcome is reported vote among validated voters; a pre-election application would replace the validation screen with a likely-voter model and could use vote intention rather than reported vote. The correction parameters are treated as fixed in the main results; the variants show that propagating their bootstrap uncertainty does not change the conclusions.

\section{Future work}

The finished 2024 pipeline can be applied to earlier elections, and the inputs exist for every presidential election since 2008, the first the CES covered. Three extensions follow. First, the nonresponse model of Section~\ref{sec:lg} requires only respondent counts, weights, and results by district, with no poststratification frame and no model fitting, so the relative response rate can be estimated for each election from 2008 to 2024; the district-level results back to 2008, on each cycle's lines, are compiled by \citet{nir2025downballot}. This would show when the online panel's differential nonresponse reversed sign relative to the telephone polls of 1988 and how it has changed since. Second, parameterizing the pipeline by election (survey variable names, ACS vintage and district geography, results files) and testing it on the 2016-to-2020 pair, then on 2008-to-2012 and 2012-to-2016, would replace one transfer with four and would allow the decay of the correction with the gap between elections to be estimated. Third, a pair of elections without a change in the coalitions can answer questions this pair cannot: whether race-specific selection is stable when the coalitions are not moving, whether the over-estimate in the most Hispanic districts recurs, and whether a citizen-voting-age frame removes it.
\newpage
\appendix

\section{Reproducibility}\label{app:repro}

Every number in this paper comes from a seeded run of the complete pipeline in a fresh R session: the 2020 backtest, then the 2024 application, then the nonresponse model and the two variants. Three input errors were caught during development by validation statistics taking impossible values, and each produced a structural guard. A fourth and a fifth, found later, are items 4 and 5.

\begin{enumerate}
\item \textbf{A shifted results file.} The 2020-results-by-2024-lines file had Arizona's block shifted by one row with a duplicated row at the bottom. It surfaced as implausible $-16$-point 2020--2024 ``swings'' for two Arizona districts in the list of out-of-band misses. After correction, Arizona's swings range from $+0.4$ to $+5.5$ points.
\item \textbf{A wrong-vintage ACS table.} A code chunk pulling race margins was lost in an edit, and the 2020 pipeline silently used a 2023-vintage object left in the session by the 2024 pipeline, joining new district lines onto old ones. The signature was eight districts with missing margins: the seven seats that no longer exist plus Montana, whose at-large identifier changed. Every ACS pull now asserts the geography it expects (Montana's at-large identifier is the sentinel), every join asserts zero missing districts, and the two pipelines are never run in one session.
\item \textbf{Unseeded sampling.} Before seeds were fixed, two realizations of the same model gave 14 and 12 out-of-band misses (12--2 and 10--2 above and below the band), against 17 (16--1) in the seeded run; coverage stayed between 96.1\% and 97.2\%, and parameter-level quantities such as $b$ moved only in the third decimal. All model fits and draw subsamples are now seeded. These runs predate the corrections of items 4 and 5; the seeded run now gives 8 misses (7--1), with MO-01, FL-23, and FL-27 inside their bands by less than 0.06 points.
\item \textbf{A frame on the previous district lines.} The 2024 frame was first built from the 2019--2023 ACS, tabulated on 118th-Congress districts. Five states (Alabama, Georgia, Louisiana, New York, and North Carolina) redrew their districts for 2024, so there the frame described different districts from the survey, the lean covariate, and the results; the Montana check of item 2 cannot detect this, since Montana has two districts on both maps. On that frame, four of the 17 districts outside the band were in Georgia and Louisiana, and the conditional Black-share coefficient of Section~\ref{sec:survives} was $+3.2$ rather than $+2.0$. It now uses the 2020--2024 ACS.
\item \textbf{Results on a newer map in one state.} North Carolina redrew its districts for the 2020 election. In the 2020 backtest the survey's district variable, the frame, and the lean covariate are on the 116th-Congress lines, but the results file was on the new map, so in the state's 13 districts estimates were compared with results for different areas. Both maps have 13 districts, so every join passed; the mismatch was found when the district lines of each input were checked after item 4. North Carolina's 2020 results are now aggregated from precinct returns \citep{vest2020}, tabulated on 2020 Census voting districts, to the 116th-Congress districts, each voting district going to the district that holds most of its area; applied to the 2016 vote, the same assignment reproduces the 2016 results by district to within 0.1 points. In the runs made at the time, the correction moved $b$ from $-0.738$ to $-0.751$ and $\lambda$ from 0.284 to 0.275, lowered the 2020 in-sample RMSE from 4.98 to 4.58, and reduced the 2024 districts outside the band from 14 to 9.
\end{enumerate}

Two details of the turnout step in Section~\ref{sec:mrp} matter for reproduction. The turnout model is the same in both years; until the final run the 2020 model lacked the state intercept and the two lean covariates, and adding them moved the 2020 district estimates by 0.04 points on average and at most 0.2, $b$ from $-0.751$ to $-0.752$, and one 2024 district (FL-23) from just outside its band to just inside. And the rescaling within state uses the total votes cast in the election being estimated: it multiplies every cell in a state by one factor, capped so that no cell exceeds full turnout, and so cancels from the district estimates wherever the cap does not bind. The cap binds in 999 of the 52,320 cells in 2024 (1.9\%) and in four in 2020.

Post-mortem regressions are implemented as guarded chunks that reproduce their stored results to numerical tolerance before computing anything new. Models are fit with \texttt{brms} 2.23.0 and Stan in R 4.4.1, four chains of 2,000 iterations; sampling takes about one minute per chain, and a complete cycle runs in roughly fifteen minutes. Code and configuration are available from the authors on request.

\section{Additional tables and figures}\label{app:tables}

\begin{figure}[H]
\centering
\includegraphics[width=0.42\textwidth]{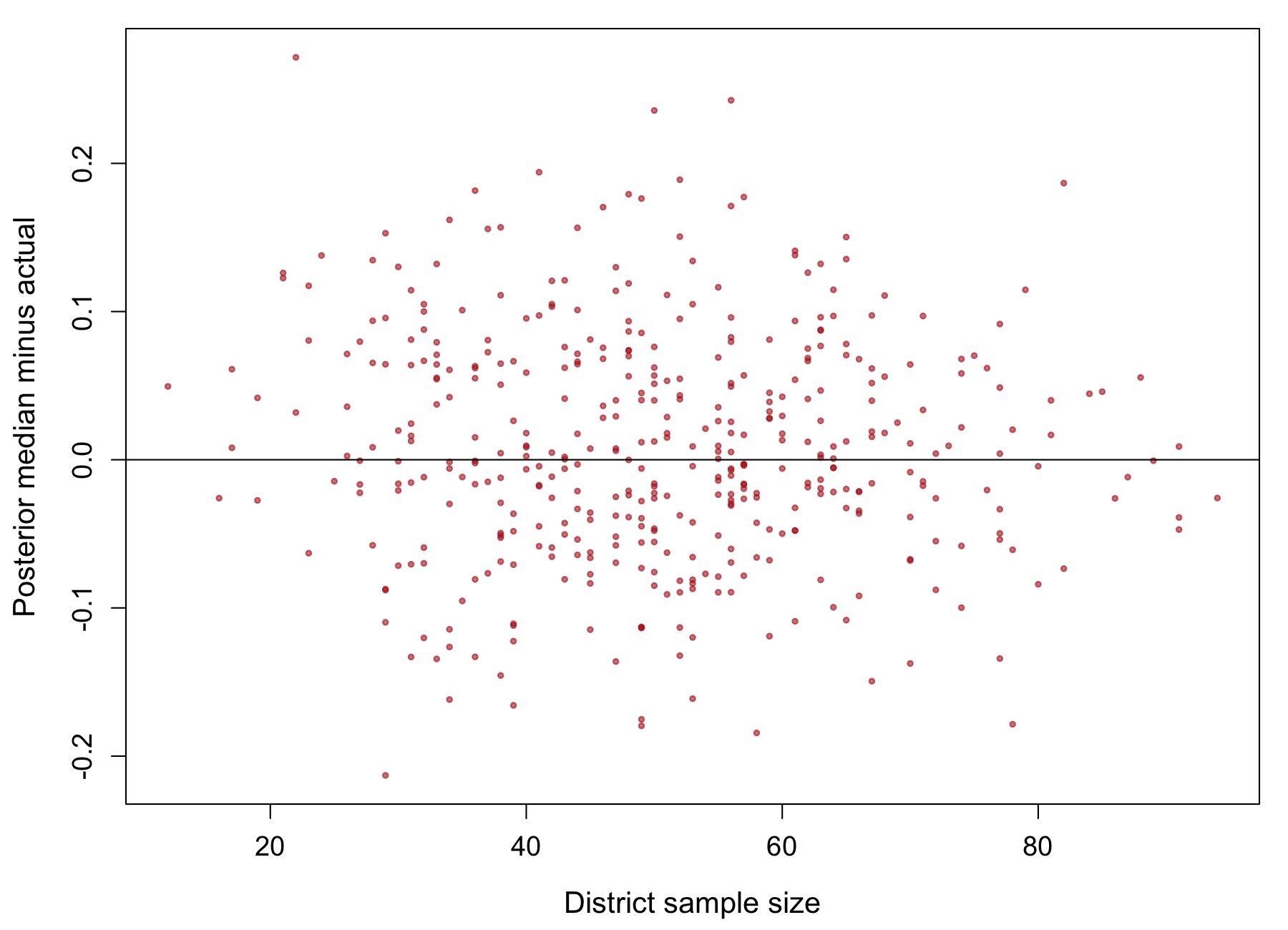}
\caption{Nonresponse model with the transferred 2020 prior: prediction error against district sample size. Errors are centered at every sample size, and the spread narrows only slightly with $m$; the remaining error comes from the exchangeable prior's inability to use demographic information rather than from small samples.}
\label{fig:errm}
\end{figure}

\begin{figure}[H]
\centering
\includegraphics[width=0.78\textwidth]{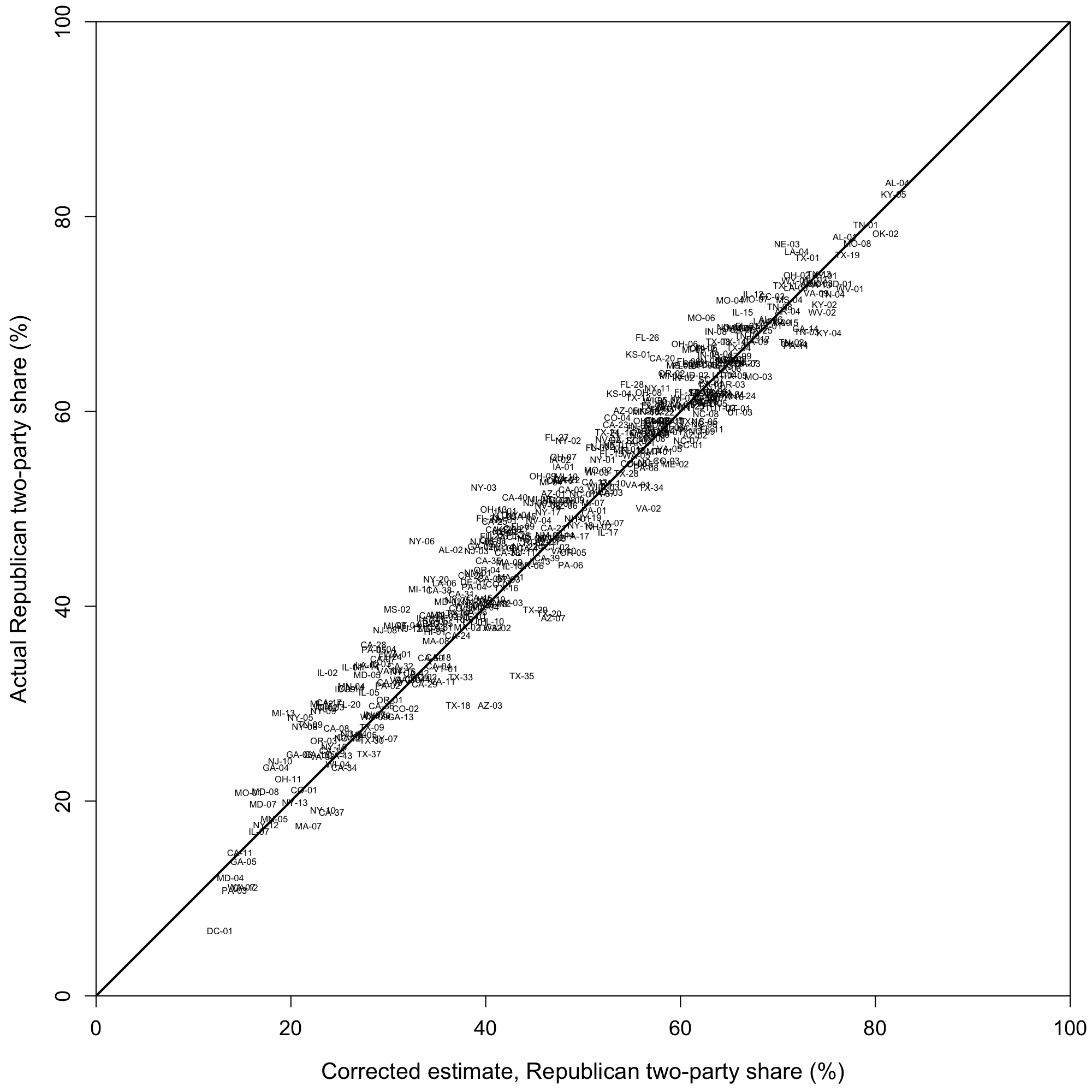}
\caption{Logit-corrected 2024 estimate against the actual result, with district labels.}
\label{fig:labeled}
\end{figure}

\begin{table}[H]
\centering\small
\caption{The 8 districts whose 2024 result falls outside the logit-corrected 95\% band, ordered by distance to the band (pp). All but DC-01 lie above their band.}
\label{tab:misses}
\begin{tabular}{lrrrrr}
\toprule
District & $n$ & Estimate & Band & Actual & Gap \\
\midrule
MI-13 & 56 & 19.3 & [12.9, 26.2] & 29.0 & 2.83 \\
NY-06 & 29 & 33.7 & [24.6, 44.9] & 46.7 & 1.86 \\
NY-03 & 34 & 40.0 & [30.1, 50.7] & 52.2 & 1.50 \\
DC-01 & 61 & 13.0 & [8.1, 19.9] & 6.7 & $-1.36$ \\
IL-02 & 37 & 23.5 & [16.1, 32.0] & 33.2 & 1.21 \\
FL-26 & 34 & 56.1 & [45.3, 66.4] & 67.6 & 1.17 \\
KS-01 & 58 & 55.2 & [44.8, 65.1] & 65.9 & 0.79 \\
AL-02 & 32 & 36.1 & [28.0, 45.3] & 45.8 & 0.52 \\
\bottomrule
\end{tabular}
\end{table}

\begin{table}[H]
\centering\small
\caption{The 44 districts in the top decile of Hispanic population share (2015--2019 ACS, 116th-Congress numbering) excluded in the robustness check of Section~\ref{sec:stagebresults}.}
\label{tab:hisp}
\begin{tabular}{p{0.9\textwidth}}
\toprule
AZ-03, AZ-07, CA-08, CA-10, CA-16, CA-20, CA-21, CA-22, CA-26, CA-29, CA-31, CA-32, CA-34, CA-35, CA-36, CA-38, CA-40, CA-41, CA-43, CA-44, CA-46, CA-51, FL-25, FL-26, FL-27, IL-04, NJ-08, NM-01, NM-02, NV-01, NY-13, NY-14, NY-15, TX-15, TX-16, TX-18, TX-20, TX-23, TX-27, TX-28, TX-29, TX-33, TX-34, TX-35 \\
\bottomrule
\end{tabular}
\end{table}
\section*{AI use}
Claude (Anthropic) was used to assist with the R code and TeX writeup of this paper. All claims, figures, and references were checked by the authors and verified against sources when necessary. The authors take full responsibility for all content, including the accuracy of the references.
\newpage
\bibliographystyle{plainnat}
\bibliography{refs}

@article{little1998modeling,
  author  = {Little, Thomas C. and Gelman, Andrew},
  title   = {Modeling Differential Nonresponse in Sample Surveys},
  journal = {Sankhy\={a}: The Indian Journal of Statistics, Series B},
  year    = {1998},
  volume  = {60},
  number  = {1},
  pages   = {101--126}
}

@article{gelman1997poststratification,
  author  = {Gelman, Andrew and Little, Thomas C.},
  title   = {Poststratification into Many Categories Using Hierarchical Logistic Regression},
  journal = {Survey Methodology},
  year    = {1997},
  volume  = {23},
  number  = {2},
  pages   = {127--135}
}

@article{park2004bayesian,
  author  = {Park, David K. and Gelman, Andrew and Bafumi, Joseph},
  title   = {Bayesian Multilevel Estimation with Poststratification: State-Level Estimates from National Polls},
  journal = {Political Analysis},
  year    = {2004},
  volume  = {12},
  number  = {4},
  pages   = {375--385}
}

@article{lax2009how,
  author  = {Lax, Jeffrey R. and Phillips, Justin H.},
  title   = {How Should We Estimate Public Opinion in the States?},
  journal = {American Journal of Political Science},
  year    = {2009},
  volume  = {53},
  number  = {1},
  pages   = {107--121}
}

@article{ghitza2013deep,
  author  = {Ghitza, Yair and Gelman, Andrew},
  title   = {Deep Interactions with {MRP}: Election Turnout and Voting Patterns Among Small Electoral Subgroups},
  journal = {American Journal of Political Science},
  year    = {2013},
  volume  = {57},
  number  = {3},
  pages   = {762--776}
}

@article{gelman2016mythical,
  author  = {Gelman, Andrew and Goel, Sharad and Rivers, Douglas and Rothschild, David},
  title   = {The Mythical Swing Voter},
  journal = {Quarterly Journal of Political Science},
  year    = {2016},
  volume  = {11},
  number  = {1},
  pages   = {103--130}
}

@article{shiranimehr2018disentangling,
  author  = {Shirani-Mehr, Houshmand and Rothschild, David and Goel, Sharad and Gelman, Andrew},
  title   = {Disentangling Bias and Variance in Election Polls},
  journal = {Journal of the American Statistical Association},
  year    = {2018},
  volume  = {113},
  number  = {522},
  pages   = {607--614}
}

@article{gelman2017nineteen,
  author  = {Gelman, Andrew and Azari, Julia},
  title   = {19 Things We Learned from the 2016 Election},
  journal = {Statistics and Public Policy},
  year    = {2017},
  volume  = {4},
  number  = {1},
  pages   = {1--10},
  doi     = {10.1080/2330443X.2017.1356775}
}

@article{heidemanns2020updated,
  author  = {Heidemanns, Merlin and Gelman, Andrew and Morris, G. Elliott},
  title   = {An Updated Dynamic {B}ayesian Forecasting Model for the {US} Presidential Election},
  journal = {Harvard Data Science Review},
  year    = {2020},
  volume  = {2},
  number  = {4}
}

@article{caughey2019public,
  author  = {Caughey, Devin and Warshaw, Christopher},
  title   = {Public Opinion in Subnational Politics},
  journal = {Journal of Politics},
  year    = {2019},
  volume  = {81},
  number  = {1},
  pages   = {352--363}
}

@article{buja2009statistical,
  author  = {Buja, Andreas and Cook, Dianne and Hofmann, Heike and Lawrence, Michael and Lee, Eun-Kyung and Swayne, Deborah F. and Wickham, Hadley},
  title   = {Statistical Inference for Exploratory Data Analysis and Model Diagnostics},
  journal = {Philosophical Transactions of the Royal Society A},
  year    = {2009},
  volume  = {367},
  number  = {1906},
  pages   = {4361--4383}
}

@article{wickham2010graphical,
  author  = {Wickham, Hadley and Cook, Dianne and Hofmann, Heike and Buja, Andreas},
  title   = {Graphical Inference for Infovis},
  journal = {IEEE Transactions on Visualization and Computer Graphics},
  year    = {2010},
  volume  = {16},
  number  = {6},
  pages   = {973--979}
}

@article{rubin1981estimation,
  author  = {Rubin, Donald B.},
  title   = {Estimation in Parallel Randomized Experiments},
  journal = {Journal of Educational Statistics},
  year    = {1981},
  volume  = {6},
  number  = {4},
  pages   = {377--401}
}

@article{burkner2017brms,
  author  = {B{\"u}rkner, Paul-Christian},
  title   = {brms: An {R} Package for {B}ayesian Multilevel Models Using {S}tan},
  journal = {Journal of Statistical Software},
  year    = {2017},
  volume  = {80},
  number  = {1},
  pages   = {1--28}
}

@article{carpenter2017stan,
  author  = {Carpenter, Bob and Gelman, Andrew and Hoffman, Matthew D. and Lee, Daniel and Goodrich, Ben and Betancourt, Michael and Brubaker, Marcus and Guo, Jiqiang and Li, Peter and Riddell, Allen},
  title   = {Stan: A Probabilistic Programming Language},
  journal = {Journal of Statistical Software},
  year    = {2017},
  volume  = {76},
  number  = {1},
  pages   = {1--32}
}

@misc{ces2020,
  author       = {Schaffner, Brian and Ansolabehere, Stephen and Luks, Sam},
  title        = {Cooperative Election Study Common Content, 2020},
  howpublished = {Harvard Dataverse, V4},
  year         = {2021},
  note         = {\url{https://doi.org/10.7910/DVN/E9N6PH}}
}

@misc{ces2024,
  author       = {Schaffner, Brian and Shih, Marissa and Ansolabehere, Stephen and Pope, Jeremy},
  title        = {Cooperative Election Study Common Content, 2024},
  howpublished = {Harvard Dataverse, V9},
  year         = {2025},
  note         = {\url{https://doi.org/10.7910/DVN/X11EP6}}
}

@misc{vest2020,
  author       = {{Voting and Election Science Team}},
  title        = {2020 Precinct-Level Election Results},
  howpublished = {Harvard Dataverse},
  year         = {2020},
  note         = {\url{https://doi.org/10.7910/DVN/K7760H}}
}

@misc{nir2025downballot,
  author       = {Nir, David},
  title        = {The {D}ownballot's Calculations of Presidential Election Results by Congressional District},
  howpublished = {The Downballot (formerly Daily Kos Elections)},
  year         = {2025},
  month        = apr,
  note         = {Published April 23, 2025; updated October 1, 2026. \url{https://www.the-downballot.com/p/the-downballots-calculations-of-presidential}}
}

@misc{downballot2024lines,
  author       = {{The Downballot}},
  title        = {2020 and 2024 Presidential Election Results by Congressional District, According to the District Lines Used in 2024},
  howpublished = {Dataset (Google Sheets), compiled as Daily Kos Elections},
  year         = {2025},
  note         = {\url{https://docs.google.com/spreadsheets/d/1ng1i_Dm_RMDnEvauH44pgE6JCUsapcuu8F2pCfeLWFo/}. Accessed February 16, 2026}
}

@misc{downballot2020lines,
  author       = {{The Downballot}},
  title        = {2008, 2012, 2016, and 2020 Presidential Election Results by Congressional District, According to the District Lines Used in 2020},
  howpublished = {Dataset (Google Sheets), compiled as Daily Kos Elections},
  year         = {2025},
  note         = {\url{https://docs.google.com/spreadsheets/d/1XbUXnI9OyfAuhP5P3vWtMuGc5UJlrhXbzZo3AwMuHtk/}. Accessed May 7, 2026}
}

@misc{downballotmethod,
  author       = {{The Downballot}},
  title        = {Methodology for Calculating Presidential Election Results by Congressional District},
  howpublished = {Google Docs, linked from \citet{nir2025downballot}},
  year         = {2025},
  note         = {\url{https://docs.google.com/document/d/1efGEHGAQGaDGnIrdedwxVg1uUrjZYz06IPLTHRCg9zM/}}
}

@article{meng2018statistical,
  author  = {Meng, Xiao-Li},
  title   = {Statistical Paradises and Paradoxes in Big Data ({I}): Law of Large Populations, Big Data Paradox, and the 2016 {US} Presidential Election},
  journal = {Annals of Applied Statistics},
  year    = {2018},
  volume  = {12},
  number  = {2},
  pages   = {685--726}
}

@article{isakov2020towards,
  author  = {Isakov, Michael and Kuriwaki, Shiro},
  title   = {Towards Principled Unskewing: Viewing 2020 Election Polls Through a Corrective Lens from 2016},
  journal = {Harvard Data Science Review},
  year    = {2020},
  volume  = {2},
  number  = {4}
}

@misc{chooi2026persistent,
  author       = {Chooi, Jay},
  title        = {The Persistent Non-Response Bias in a Sample-Matched Poll for the 2024 {U.S.} Presidential Election},
  year         = {2026},
  howpublished = {arXiv preprint arXiv:2606.12889}
}

@article{bailey2023new,
  author  = {Bailey, Michael A.},
  title   = {A New Paradigm for Polling},
  journal = {Harvard Data Science Review},
  year    = {2023},
  volume  = {5},
  number  = {3}
}

\end{document}